\documentclass[11pt,preprint,flushrt]{aastex}

\newcommand\etal{{\it et al.\/}}
\newcommand\kms{{km~s$^{-1}$}}
\newcommand\hkpc{\mbox{$h^{-1}\;\rm kpc$}}

\begin{document}
 
\slugcomment{Revision 1.7 9/29/00 gmb; submitted to ApJ 6/7/00}
 
\title{Weak Lensing Determination of the Mass in Galaxy
Halos
}
 
\author{D. R. Smith\altaffilmark{1,3}, G. M. Bernstein\altaffilmark{1}, 
P. Fischer\altaffilmark{4}, \& M. Jarvis\altaffilmark{1,2}}
\affil{Dept. of Astronomy, University of Michigan, Ann Arbor, MI 48109}
\email{deano, garyb, philf, jarvis@astro.lsa.umich.edu}
\altaffiltext{1}{Visiting Astronomer, National Optical Astronomy
Observatories, which is operated by the Association of Universities
for Research in Astronomy, Inc., under contract to the National
Science Foundation.}
\altaffiltext{2}{Based in part on research carried out at the MDM
Observatory, operated by Columbia University, Dartmouth College,
University of Michigan, and Ohio State University.}
\altaffiltext{3}{Present Address:  Science Dept, Glenelg High School,
Glenelg, MD 21737}
\altaffiltext{4}{Present Address:  Dept. of Astronomy, Univ. of
Toronto, 60 St. George St., Toronto, ON M5S3H8.}
\begin{abstract}
We detect the weak gravitational lensing distortion of 450,000
background galaxies ($20<R<23$) by 790 foreground galaxies ($R<18$)
selected from the Las Campanas Redshift Survey(LCRS).  This is the first
detection of weak lensing by field galaxies of known redshift, and as
such permits us to reconstruct the shear profile of the typical field
galaxy halo in absolute physical units (modulo
$H_0$), and to investigate the dependence of halo mass upon galaxy
luminosity.  This is also the first galaxy-galaxy lensing study for
which the calibration errors due to uncertainty in the background
galaxy redshift distribution and the seeing correction are negligible.
Within a projected radius of 200 \hkpc, the shear profile is
consistent with an isothermal profile with circular velocity
$v_c=164\pm20$ \kms\ for an $L_\ast$ galaxy, consistent with the
typical circular velocity for the disks of spirals at this luminosity.
This halo mass
normalization, combined with the halo profile derived by \citet{F00}
from a galaxy-galaxy lensing analysis of the Sloan
Digital Sky Survey, places a lower limit of
$(2.7\pm0.6)\times10^{12}h^{-1}M_\odot$ on the mass of an $L_\ast$ galaxy
halo, in good agreement with the satellite galaxy studies of \citet{Za97}
Given the known luminosity function of LCRS galaxies,
and assumption that $M\propto L^\beta$ for galaxies, we determine that
the mass within 260\hkpc\ of normal galaxies contributes
$\Omega=0.16\pm0.03$ to the density of the Universe (for $\beta=1$) or
$\Omega=0.24\pm0.06$ for $\beta=0.5$.  These lensing data suggest
that $0.6<\beta<2.4$ (95\% CL), only marginally in agreement with the
usual $\beta\approx0.5$ Faber-Jackson or Tully-Fisher scaling.
This is the most complete direct inventory of the matter content of the
Universe to date.
\end{abstract}


\section{Introduction}
Gravitational lensing provides the only non-dynamical measure of the
mass distribution of a galaxy halo, and it thus enables a crucial
means of quantifying the dark matter halos that are now an essential
component of all theories of cosmology and galaxy formation.  A
reliable measure of the masses of galaxies and their halos also
provides a significant lower limit to the density parameter $\Omega$.
Spiral galaxy halos have traditionally been probed via rotation of
stars and gas, but these tracers are not available outside radii of
$\sim30$~kpc, though their velocities are consistent with with
an isothermal halo extending beyond this distance.  At larger radii,
dynamical test particles are scarce: \citet{Za97} use
satellite galaxies to trace the halo mass profile to $\approx200$~kpc.
Because there are only $\approx1.2$ satellites per spiral, the masses
of individual galaxies are poorly constrained, but by ``stacking'' the
signals from an ensemble of galaxies, an accurate measure of the mean
halo is obtained.  A further complication of such dynamical studies of
the outer halo is that the test particles are not virialized, and
determination of the halo mass requires some model of
infall.  With such modelling, \citet{Za97} estimate a
mass of $\sim2\times10^{12}\,M_\odot$ within 200~kpc of an $L_\ast$ spiral
galaxy.

The gravitational lensing approach uses photons from background
galaxies (bggs) as the test particles, enabling halo measurements to
large radii.  The greatest problem with 
lensing is the extremely weak shear ($\lesssim3\%$) associated with
the mass of a single galaxy.  This is more than a factor of 10 below
the noise due to intrinsic shape variations in a single background
galaxy.  As with the satellite galaxy study, the weak lensing approach
must therefore overcome the poor $S/N$ per galaxy by using large
numbers of lens-source pairs, measuring the average galaxy halo. The
weak lensing measurement, however, requires no modeling of the
dynamical state and can give a non-parametric estimate of the halo
profile.  The first attempted detection was \citet{Ty84}, and
over the last five years several groups have verified the
existence of galaxy-galaxy lensing based upon analysis of a small
number of deep fields \citep{Br96, DA96, Gr96, Hu98}.
Interpretation of these results is problematic not only
because of poor $S/N$, but also because the redshifts of the lens
galaxies (and hence their distances and luminosities) are not known
individually.  Interpretation relies upon positing a model for the
joint distance/luminosity distribution of the lenses and sources, and
a model for dependence of halo mass and size upon luminosity, then
projecting these models down to a function of angular variables to be
compared with the observations.  In such deep fields, even the
statistical distributions of lens and source distances are currently
ill-determined.

More recently \citet[F00]{F00} have measured a galaxy-galaxy lensing
signal at high significance from preliminary Sloan Digital Sky Survey
data.  These data are relatively shallow (foreground galaxies
$16<r'<18$, background galaxies $18<r'<22$), which weakens the shear
signal, but this is more than compensated by the extremely large
sample size of $\approx16$ million fgg/bgg pairs. These data indicate
that halos of typical galaxies continue an isothermal profile to a
radius of at least 260~\hkpc.  The conversion from shear amplitude to
mass density in the F00 data is still reliant upon photometric
redshift distributions for the fggs and bggs.

We present here a detection of galaxy-galaxy lensing around fggs with
redshifts determined by the Las Campanas Redshift Survey \citep{Sc96}
We thus have knowledge of the fgg luminosity and
distance, and the impact parameter of the lensed photons, in physical
units ({\it i.e.\/} kpc rather than arcsec), information that is
normally lost in projection.  The bgg's
($R<23$) are at magnitudes within the reach of current
pencil-beam redshift surveys, and hence their distance distribution is well
enough determined to introduce negligible uncertainty in the
calibration.  We are therefore in theory able to measure the galaxy
luminosity $L$, halo mass $M$, and surface mass density $\Sigma$ in
physical units without ambiguity (save the scale factor $H_0$).  In
practice our $S/N$ is too low to constrain the extent of the galaxy
halo meaningfully, so we combine the F00 profile {\it shape} with our
{\it normalization} to obtain a mass profile in absolute units.
Combining the resultant $M/L$ with the well-determined luminosity
function of the LCRS galaxies \citep{Li96} we obtain the galactic
contribution to $\Omega$.  
We assume an Einstein-deSitter
Universe for the values presented here, and use $H_0=100h {\rm \,km\,
s}^{-1}\,{\rm Mpc}^{-1}$.  Because the fggs are relatively nearby, and
the bggs are far behind them, the
dependence of the results upon the cosmological parameters $\Omega$
and $\Lambda$ is negligible ($<2\%$).

\section{Observations and Reduction Methods}

\subsection{Observations}
We observed 36 square degrees of sky in $R$ band using the Big
Throughput Camera (BTC) \citep{Wi98} on the CTIO Blanco
4-meter telescope over three runs from December 1996 through March
1999.  Most areas of the target fields are covered by three dithered
420-second exposures.  Observations were made within the Las Campanas
Redshift Survey fields and yield $\approx45,000$ well-measured bggs
per ${\rm\,deg}^2$, with typical seeing of 1\farcs1 FWHM and RMS sky noise
of $\approx28\, R {\rm \,mag\,arcsec}^{-2}$ on 0\farcs43 pixels.  Images are
debiased and flattened using standard IRAF\footnote{
IRAF is distributed by the National Optical Astronomy Observatories,
which is operated by the Association of Universities for Research in
Astronomy, Inc. (AURA) under cooperative agreement with the National
Science Foundation.}
routines.  Rather than sum images, we analyze each 300s exposure
separately and average the resulting galaxy measurements.  This allows
us to adjust for exposure-to-exposure seeing variations as well as to
avoid or isolate various systematic errors.  Observations of $R$-band
standards from \citet{La92} yield photometric solutions with
worst-case RMS errors of 0.05~mag.

\subsection{Shape Measurement}
Galaxy shape measurements are critical to weak lensing.  Detailed
descriptions of our methods are presented in \citet{Be00} and
\citet{Sm00}. Here we summarize the processing steps.  Initial
detection, photometry, and size estimation is done by {\tt SExtractor}
\citep{Be96}.  This information is then fed to our shape
measurement code {\tt elliptomatic}, which determines ellipticities
from Gaussian-weighted second moments of each object.  The Gaussian
weights are elliptical; the weight ellipticity is iterated to match
the object's ellipticity, which produces an unbiased estimate of the
object shape.  Stellar objects are selected from the catalog, and a
position-dependent convolution kernel is created which, when applied
to the original image, makes the stellar PSF round everywhere in the
image \citep{Fi97, Be00}
This corrects all objects for shape errors induced by
telescope tracking, seeing, or linear CCD charge transfer inefficiencies.
Optical distortions are significant across the wide field of the BTC.
Observations of astrometric standard fields yield a distortion map,
and object shapes are corrected for this distortion analytically,
yielding the observed (post-seeing) ellipticity components $e_1^o$ and
$e_2^o$.  The observed shapes are then corrected for the circularizing
effects of the PSF to give an estimate of the pre-seeing ellipticity
$e_1^i$ and $e_2^i$ which would be observed in an image with perfect
resolution.  If this correction is larger than a factor of 5, the
object is discarded.  The measured ellipticities of a given
object on different exposures are averaged to give the mean pre-seeing
ellipticity for that object.  An estimated uncertainty $\sigma_e$ due
to image noise is also calculated for each object.

When considering the distortion induced by a given fgg, the
bgg ellipticity is rotated to components $e_+$ and $e_\times$, where the
former is positive for an image oriented tangentially to the fgg.

The algorithms used herein to correct for PSF anisotropy and
circularization are known to leave systematic residuals at the
$\sim0.5\%$ level in the shapes of the galaxies.  These have little
effect upon the galaxy-galaxy lensing measurements since one is
measuring $\langle e_+ \rangle$ in annuli about the fggs, and a
systematic distortion of constant orientation will cancel to first order when
integrated around the annulus.  We will demonstrate the absence of
significant systematic errors below.

\subsection{Determination of Distortion from Ellipticities}
A given bgg with source-plane tangential ellipticity $e_+^s$, when
subjected to a weak (tangential) distortion $\delta$, will in the
absence of noise be measured to have image-plane ellipticity
\begin{equation}
e_+^i = e_+^s + \delta [1-(e_+^s)^2].
\end{equation}
Since the bgg orientations are assumed isotropic, $\langle e_+^s
\rangle_b = 0$, 
a useful unbiased estimator for the distortion is
\begin{equation}
\hat\delta = \langle e_+^i \rangle_b / {\cal R}, 
	\qquad {\cal R}\equiv \langle 1-(e_+^i)^2 \rangle_b
	\equiv 1-\sigma^2_{SN},
\end{equation}
where $\sigma^2_{SN}\equiv \langle (e_+^i)^2 \rangle_b$ is often
called the {\it shape noise}.
The {\em responsivity} $\cal R$ is similar to the mean ``shear
polarizability'' of \citet{Ka95}.
In the presence of measurement error $\sigma_e$ upon each ellipticity,
we may wish to apply some weight function $w(e,\sigma_e)$ to each bgg
shape.  The estimator for local distortion becomes \citep{Be00}
\begin{eqnarray}
\label{hatd}
\hat\delta & = & {1 \over {\cal R}} {\sum_b w e_+^i \over \sum_b w}
\\
\label{resp}
{\cal R} & \equiv & \sum_b \left[w(1-fe_+^2) + 
	{\partial w \over \partial e}{e_+^2 \over e}(1-fe^2)
	\right] \over \sum_b w \\
f & \equiv & \sigma^2_{SN} \over \sigma^2_{SN}+\sigma^2_e.
\end{eqnarray}
The parameter $f$ describes the fraction of the variance in measured
$e_+^i$ for the bgg population which is attributable to shape noise.
Any weight function is permissible if it depends only upon the
magnitude, not the direction, of the ellipticity $e$.  We use the
following weight, which can be shown to minimize the variance of the
estimator under certain conditions \citep{Be00}:
\begin{equation}
\label{wt1}
w(e,\sigma_e) = {1 - (fe^o)^2 + \sigma^2_{SN}(4f-1)
	\over \sigma^2_{SN}+\sigma^2_e }.
\end{equation}
Alternative choices of weight function lead to results that are
indistinguishable within the noise.

\begin{figure}
\plotone{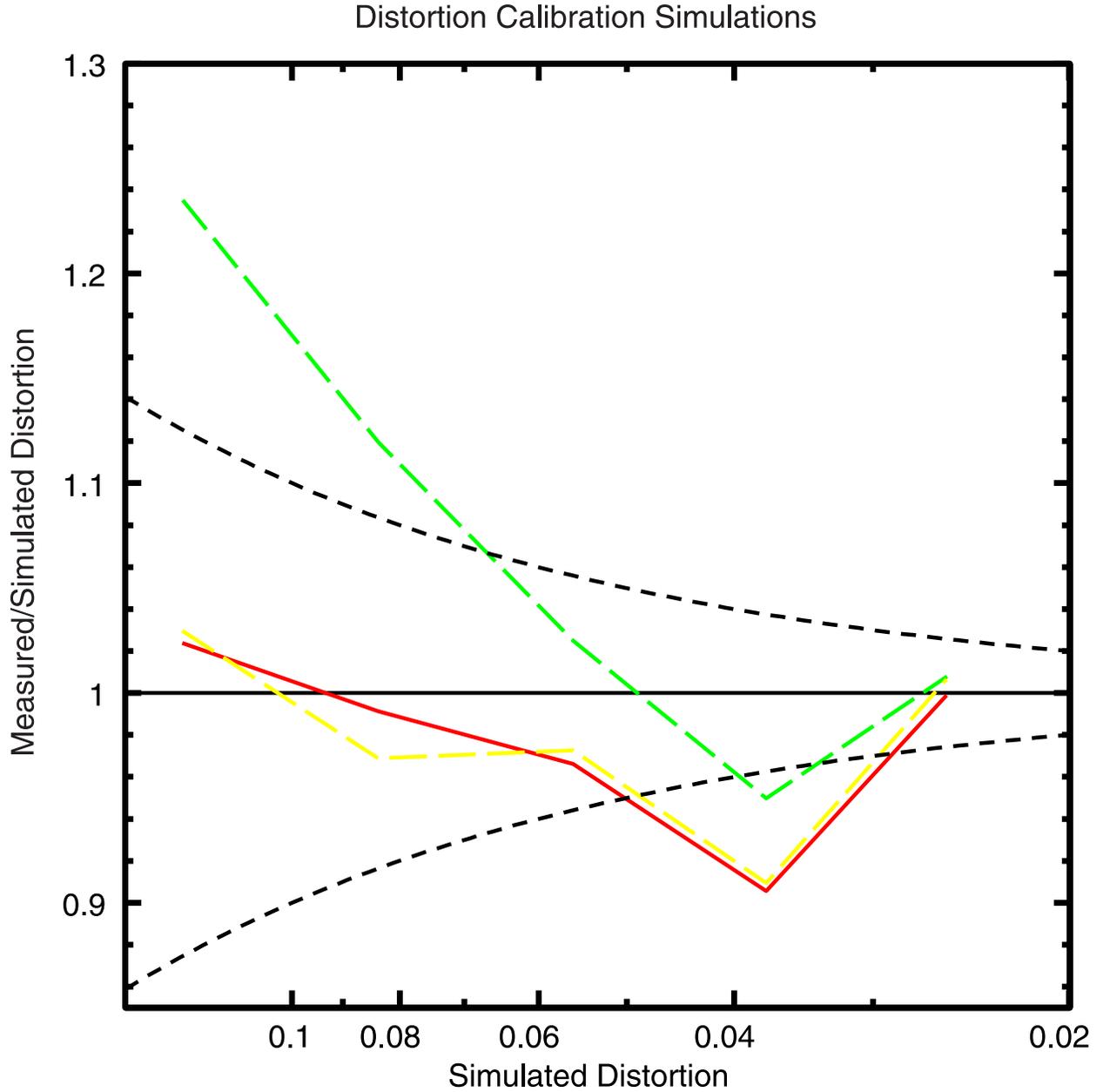}
\caption[dummy]{\small
Results of shear calibration with simulated data:  a simulated
isothermal distortion
is applied to artificial galaxy images, which are then measured via
the standard measurement pipeline.  The ratio of the measured to the
applied distortion is shown for our standard weighting scheme (solid
line) and two alternative weighting schemes (dashed lines).  
Ideally the ratio is unity to order $e$, {\it i.e.\/} will fall
between the two dashed curves.
The measured values are within $\sim5\%$ of the applied distortion in
the limit of weak lensing, and are nearly independent of weighting
scheme.  The results are not visibly changed when the artificial
galaxy sizes are reduced by 30\%.
\label{calsims}
}
\end{figure}

We have tested the accuracy of the distortion-measurement algorithms
using simulated data frames.  An image of artificial spiral and
elliptical galaxies is distorted by a circular isothermal potential,
then subjected to degradation by Gaussian seeing and noise similar to those in
our images.  The artificial images are then run through the software
pipeline, and the above equations used to extract the distortion
$\delta$ as a function of radius from the putative lens.  The results
are plotted in Figure~\ref{calsims}, which illustrates that the input
distortion is recovered to an accuracy of 5\% or better as long as the
distortion is indeed weak.  To insure that the calibration of PSF
circularization is correct and robust, we repeat the simulation with
the angular sizes of the bgg's arbitrarily reduced by 30\%.  The input
distortion is again recovered to the same accuracy.

An additional complication of the real data is that the bgg's are
distributed in redshift but the distances to individual galaxies are
not known.  Pencil-beam redshift surveys give the joint distribution
$N(m,z)$ of the background population.  We will assume that
the responsivity $\cal R$ and weight $w$ of the bgg's are
uncorrelated with bgg redshift $z_b$ at a given bgg magnitude $m_b$.
Then the
$\hat\delta$ quantity in Equation~(\ref{hatd}) is a scaled estimator
of the distortion $\delta_\infty$ that would be measured for a bgg
population at infinite redshift:
\begin{eqnarray}
\label{dinf}
\langle \hat\delta \rangle & = & \delta_\infty
	{ \sum_b w \bar F_D(z_f,m_b) \over \sum_b w} \\
\label{fd} 
\bar F_D(z_f, m_b) & = & { \int_{z_f}^\infty (D_{fb}/D_{ob})N(m_b,z_b)\,dz_b
	\over \int_0^\infty N(m_b,z_b)\,dz_b}.
\end{eqnarray}
The $D$'s are angular diameter distances between observer, foreground,
and background redshifts.
The quantity $\delta_\infty$ is thus measurable as long as the
distribution of the background weight vs redshift is known.  The
redshift distribution of our bgg sample is sufficiently well known
from deep redshift surveys that the uncertainty in the factor $\bar
F_D$ does not dominate the error budget, as will be discussed in
detail below.

\subsection{Mass Density and Scaled Distortion}
The physical quantity of interest is the mean azimuthally averaged surface
mass density $\Sigma_L(R)$ at radius $R$ about galaxies of luminosity $L$.
The measured quantity is the lensing distortion $\delta_\infty$ induced upon
a distant bgg $z_b$, which is related to the surface density by
\begin{equation}
\label{dist1}
\delta_\infty(R,L,z_f) = \left[ \Sigma_L(\le R) - \Sigma_L(R) \right]
 { 8\pi G \over c^2} D_{of}
\end{equation}
Here $\Sigma_L(\le R)$ is the average surface mass density interior
to $R$.
We will wish to combine data from fggs which 
span a range of $z_f$, and we additionally will bin data over finite
ranges in impact
parameter $R$, and fgg luminosity $L$.  To
compensate for the spread in these quantities, we
express the distortion $\delta_\infty(R,L,z_f)$ for a given fg/bg pair
in terms of a scaled distortion
\begin{eqnarray}
\label{dstar}
\delta_\ast & \equiv & \delta(R=R_0,L=L_\ast,z_f=z_\ast, z_b=\infty) \\
& = & \delta_\infty(R,L,z_f)  / \left[F_L(L) F_R(R) F_z(z_f)
\right].
\end{eqnarray}
Our canonical luminosity is the $L_\ast$ in the Schechter-function
parameterization of the LCRS galaxy luminosity function, which is
absolute $R$-band magnitude $M_\ast=-20.3+5\log h$, or
$L_\ast=8.4h^{-2}\times10^{9} L_\odot$ \citep{Li96}
$z_\ast$ is taken to be $0.1$, the median redshift of the LCRS sample;
and $R_0$ is taken as 100\hkpc\ if not otherwise specified as the
center of some radial bin.
The scaling factor for redshift is known:
\begin{equation}
\label{fz}
F_z(z_f)  =  D_{of}(z_f) / D_{of}(z_\ast)
\end{equation}
Scaling the measured shear to a standard luminosity and radius
requires some model for how the surface density profile varies with
these quantities.  With a sufficiently large galaxy sample, one could
simply measure the function
$\Sigma_L(\le R)-\Sigma_L(R)$ in narrow bins of $L$ and $R$, and
non-parametrically map this density contrast as a function of these two
variables.  Our samples are, however, small enough that our bins must cover a
substantial range of $L$ and $R$, so we will adopt the
assumption that the surface density is described by an isothermal
profile with a circular velocity $v_c$ that is a power-law function of
luminosity:
\begin{equation}
\label{vcbeta}
\Sigma_L(R) = \Sigma_L(\le R)/2  =  {v_c^2(L) \over
4GR}, \qquad 
v_c(L)  =  v_\ast(L/L_\ast)^{\beta/2}.
\end{equation}
The scaling functions are then very simple:
\begin{equation}
\label{flr}
F_L  =  (L/L_\ast)^{\beta}, \qquad 
F_R  =  R_0/R.
\end{equation}
The canonical distortion is
\begin{equation}
\label{dstar2}
\delta_\ast  =  2\pi {v_\ast^2 \over c^2} {D_{0f}(z_\ast) \over R_0}
	=  0.0071 \left({v_\ast \over 200\,{\rm km}\,{\rm s}^{-1}}\right)^2
\left({R_0 \over 100\hkpc}\right)^{-1}
\end{equation}

The high-S/N data of F00 produce a mean distortion profile that is
fully consistent with the isothermal profile at the radii
$R\lesssim200\hkpc$ in which we measure $\delta$.  The power-law
dependence of $\Sigma$ upon galaxy luminosity $L$ is suggested by the
Tully-Fisher and Faber-Jackson relations, which lead us to choose the
value $\beta=0.5$.  These relations have, however, been tested only at
radii of $\sim30\hkpc$ or smaller---the F00 study cannot test this
since the fgg luminosities are not known individually.  Indeed
\citet{Za97} were unable to detect {\it any} dependence of
halo mass upon luminosity in the spiral satellite study.  We will also
calculate $\delta_\ast$ under the assumption $\beta=1$, the
mass-traces-light value.  In \S\ref{lscaling} we will use our own data
to bound $\beta$.

The goal of
this paper is an absolute determination of  $\delta_\ast$.
Our estimator for $\delta_\ast$ combines
Equations~(\ref{hatd}), (\ref{dinf}), and (\ref{dstar}).  In addition
the variance of the estimator is minimized by adding a factor of
$(F_L F_R F_z \bar F_D)^2$ to the weight function.  We thus obtain
\begin{equation}
\hat\delta_\ast = {1 \over \cal R} { \sum_{f,b} w F_L F_R F_z \bar
F_D e_+^i \over  \sum_{f,b} w [F_L F_R F_z \bar F_D]^2}
\end{equation}
Standard propagation of errors, along with the substitution
${\rm Var}(e_+)=\langle e_+^2 \rangle$, yields the variance of the
above estimate:
\begin{equation}
\label{vard}
{\rm Var}(\hat\delta_\ast) = 
	{ \sum_{f,b} \left[ w F_L F_R F_z \bar F_D e_+^i\right]^2
    \over  {\cal R}^2 \left\{ \sum_{f,b} w [F_L F_R F_z \bar F_D]^2\right\}^2}
\end{equation}

\section{Results}
Our foreground sample consists of all 790 galaxies from the LCRS redshift
survey with $0.05<z_f<0.167$ that lie within our imaged areas.
Background galaxies are required to be well-measured on at least two
exposures, have magnitude $20<R<23$, 
and correction for PSF circularization at most a factor 5 (this is
effectively a lower limit on angular size).  This leaves $\approx450,000$
fgg/bgg pairs with impact parameters of $15\hkpc\le R\le480\hkpc$.  

\subsection{Background Galaxy Redshift Distribution}
An absolute determination of mass densities requires an estimate of
the distance factor $\bar F_D(z_f,m_b)$ from the weighted redshift distribution
of the background galaxies.  The Caltech Redshift Survey of the Hubble
Deep Field and its flanking fields \citep[CRS]{Co00} provides a
virtually complete redshift survey to $R\le23$.
We combine the CRS $N(m,z)$ data into 0.5-mag bins to construct the
function $\bar F_D$.  This factor is above 0.6 for nearly all the
fgg/bgg pairs in our data, and is by definition $\le1$.  This limited
range of variation is a consequence of the substantial magnitude
(hence distance) gap between our foreground and background samples,
and makes our calibration very stable against perturbations in the
assumed $N(m,z)$.  Thus while the HDF field may be deficient in nearby
galaxies, we do not expect a significant effect upon our results.

To estimate the uncertainty in $\delta_\ast$ due to $\bar F_D$ errors we
turn to the Canada-France
Redshift Survery \citep[CFRS]{Cr95}, which is nearly complete for a
sample of $\approx600$ galaxies in 5 fields defined by $I_{AB}<22.5$. We have
imaged two of these fields in $R$ band using the 2.4-meter Hiltner
telescope at MDM Observatory to obtain an $N(m_R,z)$ distribution of
the CFRS sample.  We can either use the 
measured $N(m_R,z)$ in these two fields, or use them to construct a
mean $(R-I_{AB})$ vs $(V_{AB}-I_{AB})$ color relation for the
galaxies, and apply this relation to the full CFRS catalog to mimic an
$R$-band observation.  The values of $\delta_\ast$ determined by the
two methods differ by only $\approx1\%$.  

Using the measured CFRS $N(m,z)$ changes $\delta_\ast$ by $\pm5\%$ or
less relative to the use of the CRS.  The CFRS is substantially
incomplete in two senses:  first, the $I_{AB}<22.5$ cutoff of the CFRS
catalog misses an unknown number of the bluer $R<23$ galaxies.  These
bluer galaxies are likely nearer, on average, then the measured $R\approx23$
sample.  Second, the CFRS redshift measurements are incomplete even
for the  $I_{AB}<22.5$ sample; this incompleteness is as large as 30\%
in the $22.5<R<23.0$ bin.  These galaxies are likely to be biased
toward the $1<z<2$ range where redshift measurement is difficult.  If
we place all the unmeasured galaxies at $z=2$, $\delta_\ast$ is still
only 6\% lower than the CRS result---and it is likely that inclusion
of the missing blue $R\approx23$ galaxies would push this number back
toward the CRS result.

We conclude that calibration errors due to uncertainties in the bgg
redshifts are $<5\%$.

\subsection{Measured Halo Properties}
\begin{figure}
\plotone{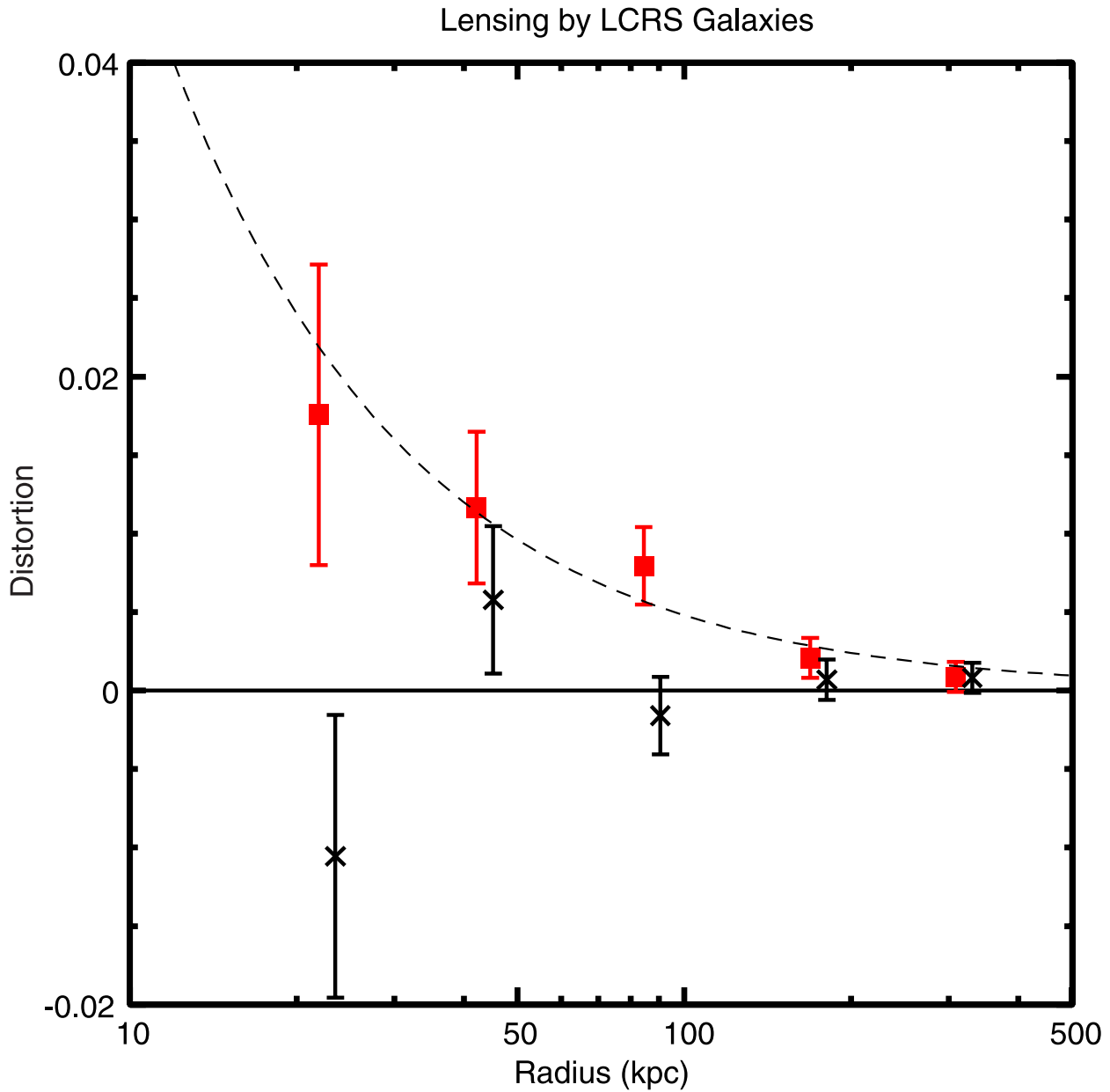}
\caption[dummy]{\small
The square symbols show the measured
distortion $\delta_\ast$ (scaled to an $L_\ast$ galaxy at $z_f=0.1$
and $z_b=\infty$) vs impact parameter.  The dashed line is the
best-fit isothermal model.  The cross symbols ($\times$) show the
binned means of the $\delta_\times$ component of shear, which should
(and does) vanish if the distortions are due entirely to lensing.
This plot depicts results assuming $M\propto L^\beta$ with $\beta=1$,
and the total significance of the detection is $4.5\sigma$.  The
results for $\beta=0.5$ are similar.
\label{distvsr}
}
\end{figure}

We plot the scaled distortion $\delta_\ast$ as a function of proper
distance from the lens galaxy center in Figure~\ref{distvsr}.  The
tangential alignment is clearly detected to $\sim150\hkpc$.  As a test
for systematic contamination, we calculate the azimuthal average of
$e_\times$.  The $e_\times$ averages are consistent with
zero distortion, as they must be if the distortion is due to weak
lensing \citep{St96}, 
giving us confidence that systematic shape distortions have
minimal effect upon our $\delta_\ast$ determination.

Our data are consistent with an isothermal profile but are too noisy
to provide any useful constraint on the profile shape.  No useful
lower limit, for example, is implied upon a possible outer truncation
radius $R_{\rm max}$ for an isothermal halo.  The high-S/N data of
F00, however, do provide useful constraints to the truncation radius
which we will use below.

The most accurate determination of $\delta_\ast$ is made by fitting
all the data with $15\hkpc<R<240\hkpc$ to the isothermal model. 
For the two trial values of $\beta$ we obtain
\begin{equation}
\label{measured}
\delta_\ast = \cases{
0.0056\pm0.0014 & $\beta=0.5$ \cr
0.0048\pm0.0010 & $\beta=1.0$ \cr}
\end{equation}
In either redshift model the lensing signal is detected at
signal-to-noise ratio of 4--5.  For comparison, the $\delta_\times$
values computed over this broad bin are $(1.\pm11)\times10^{-4}$ ($\beta=1$), 
consistent with zero within the calculated uncertainties.

\subsection{Sources of Error}
Possible sources of error in our determination of $\delta_\ast$
include:
\begin{enumerate}
\item {\it Noise:} the largest source of error is the random shape
noise and measurement noise in the ellipticities of the bggs.  This
contributes an uncertainty of 20--25\% (1$\sigma$) to
$\delta_\ast$, and is the error quoted in Equation~(\ref{measured}).   

\item {\it Background Redshifts:} As discussed above, the $R<23$
redshift distribution is known well enough to reduce calibration
uncertainty to $\pm5\%$ or less.

\item {\it Distortion Calibration:} The accuracy with which our
software measures the galaxy shapes, corrects to the pre-seeing
shapes, and calculates a distortion has been tested by the 
simulations.  This accuracy appears to be $\approx5\%$ or better.

\item {\it Mis-measurement:} One could imagine that the shape
measurements of background galaxies could be biased by the
wings of the foregound galaxies.  At radii
beyond 30\hkpc, however, the light from the foreground galaxies is
far below the noise of the images.

\item {\it Luminosity Scaling:} The measured $\delta_\ast$ depends to
some degree on the index $\beta$ scaling luminosity to mass
[Equation~(\ref{vcbeta})].  We will attempt to measure $\beta$ below.

\item {\it Satellite Galaxies:} Our sample of ``background galaxies'' 
will be diluted by faint galaxies that are physically associated with
the foreground galaxies.  $\delta_\ast$ should be adjusted upward by
a factor equal to the fraction of bgg's that are in fact satellite
galaxies.  F00 measure this contamination to be about 15\% at
a projected radius of $\sim60\hkpc$, where our data are centered.
Our fgg sample is similar to that of F00, but the bgg magnitude ranges
differ:  $18<r^\prime<22$ for F00, and $20<R<23$ for us.  Our fainter
bgg sample should have a substantially smaller fraction of physically
associated galaxies since background counts rise much more quickly
(factor of 2.5 per magnitude in $R$, \citep{Ty88})
than the dwarf galaxy luminosity function (factor of 1--1.5 per mag).  
We detect this effect in our own data, but it is difficult to measure
precisely. We estimate, however, that our
contamination factor should be below the 10\% level.

\item {\it Projected Neighbors:} The distortion $\delta(R)$ measures
the surface mass density (above the cosmic mean density) projected
within radius $R$ of a fgg.  As galaxies are clustered in space, a
circle of radius $R$ about a fgg actually includes mass from the halos
of excess neighboring galaxies.  The strength of this effect can be
estimated by convolving the galaxy-galaxy correlation function
$\xi(r)\approx(r/8h^{-1}{\rm Mpc})^{-1.8}$ with an isothermal mass
distribution.
At our radii of $R\sim100\hkpc$ the effect is quite small ($\approx4\%$)
and can be ignored.  Beyond 200\hkpc\ the effect rises above 10\% and
must be taken into account (see F00).
Note that this effect opposes that of satellite
``dilution'' mentioned previously, but both tend to flatten the run of
$\delta$ with radius.

\end{enumerate}
In summary, the errors in $\delta_\ast$ are dominated by random noise
at $\pm20\%$.  There are 4 calibration uncertainties each at
$\lesssim5\%$ level, which are known to work in dissimilar directions,
so the overall calibration uncertainty should be
below 10\% (save the uncertainty in $\beta$), and will be ignored.

\subsection{Total Mass in Galaxy Halos}
The halo mass profiles are consistent with an isothermal profile to
the $\sim200\hkpc$ radii we probe; the total mass associated with each
galaxy depends upon the extent of these isothermal halos.
If the surface density of a galaxy follows the isothermal profile in
Equation~(\ref{vcbeta}) to a projected radius $R_{\rm max}$, beyond
which it is abruptly truncated, then the total mass of a typical
galaxy at luminosity $L$ is
\begin{eqnarray}
\label{mtot1}
M(L) & = & M_\ast (L/L_\ast)^\beta, \\
M_\ast & = & {\pi \over 2G} v_\ast^2 R_{\rm max} \\
 & = & 2.06\times10^{14}h^{-1}M_\odot \,\delta_\ast
	\left({R_{\rm max} \over 100\hkpc}\right).
\label{mtot2}
\end{eqnarray}
Our data do not constrain $R_{\rm max}$, so we turn to the SDSS data.
The F00 distortions are fit to a model in which all galaxies
are truncated at some angular size $s$;
F00 obtain a 95\% CL lower limit of $s>140\arcsec$ for their
sample.\footnote{
The fitting function in F00 is not quite a cylindrical truncation at
$s$, but the integrated mass is the same.}
The estimated redshift distribution of fggs in the F00 sample
has a weighted mean angular diameter distance of $0.125c/H_0$, so a
simple estimate of the physical scale of the truncation radius is
$s>260\hkpc$ at 95\% CL.  The F00 model ignores possible variations of
halo extent with galaxy luminosity, but such considerations are
not likely to influence our result because the mass integrals and the
F00 data are both dominated by galaxies near $L_\ast$.  We will
therefore adopt from F00 the result
\begin{equation}
R_{\rm max}\ge260\hkpc \qquad (95\%\,{\rm CL}).
\end{equation}

Taken with Equations~(\ref{measured}) and (\ref{mtot2}), 
this implies that the total mass and
mass-to-light ratio for an $L_\ast$ galaxy contained within projected
radius of 260\hkpc\ are:
\begin{eqnarray}
M(<260\hkpc) & = & 
\cases{
(3.1\pm0.8)\times10^{12}\,h^{-1} M_\odot & $(\beta=0.5)$ \cr
(2.7\pm0.6)\times10^{12}\,h^{-1} M_\odot & $(\beta=1.0)$ \cr
} \\
\label{ml}
M(<260\hkpc)/L & = & 
\cases{
(360\pm90)h\,M_\odot / L_{\odot,R} & $(\beta=0.5)$ \cr
(310\pm60)h\,M_\odot / L_{\odot,R} & $(\beta=1.0)$ \cr
}
\end{eqnarray}
Note that the LCRS luminosities are isophotal, and \citet{Li96}
recommend raising the luminosity estimates (hence lowering $M/L$) to
account for flux excluded from the apertures.  There has also been no
account taken of evolution since $z\sim0.1$, though a $K$-correction
is made.

Using the Schechter-function parametric fit to the LCRS galaxy
luminosity function from \citep{Li96} and the mass formula in
Equation~(\ref{mtot1}), the mean mass density in galaxy halos is
$M_\ast \phi_\ast \Gamma(1+\alpha+\beta)$, where
$\phi_\ast=0.019h^{3}{\rm Mpc}^{-3}$ and $\alpha=-0.70$ are the fitted
Schechter-function parameters.  As a fraction of the critical density, the
density in LCRS halos is
\begin{equation}
\label{omega}
\Omega_{\rm halo} = \cases{
16.4\;\delta_\ast \left({R_{\rm max} \over 100h^{-1}\,{\rm kpc}}\right)
 \ge 0.23\pm0.06 & $(\beta=0.5)$ \cr
12.7\,\delta_\ast \left({R_{\rm max} \over 100h^{-1}\,{\rm kpc}}\right)
 \ge 0.16\pm0.03 & $(\beta=1.0)$ \cr
}
\end{equation}
These constitute a lower limit to $\Omega$ since the matter associated
with galaxies may well extend beyond 260\hkpc.  Furthermore we have
only inventoried matter which is associated with galaxies included in
the LCRS sample.  Low-surface-brightness galaxies excluded
from the LCRS would increase $\Omega$ to some extent, as of course
would any form of matter which is not spatially correlated with the
centers of LCRS galaxies.

\subsection{Mass vs Luminosity}
\label{lscaling}
\begin{figure}
\plotone{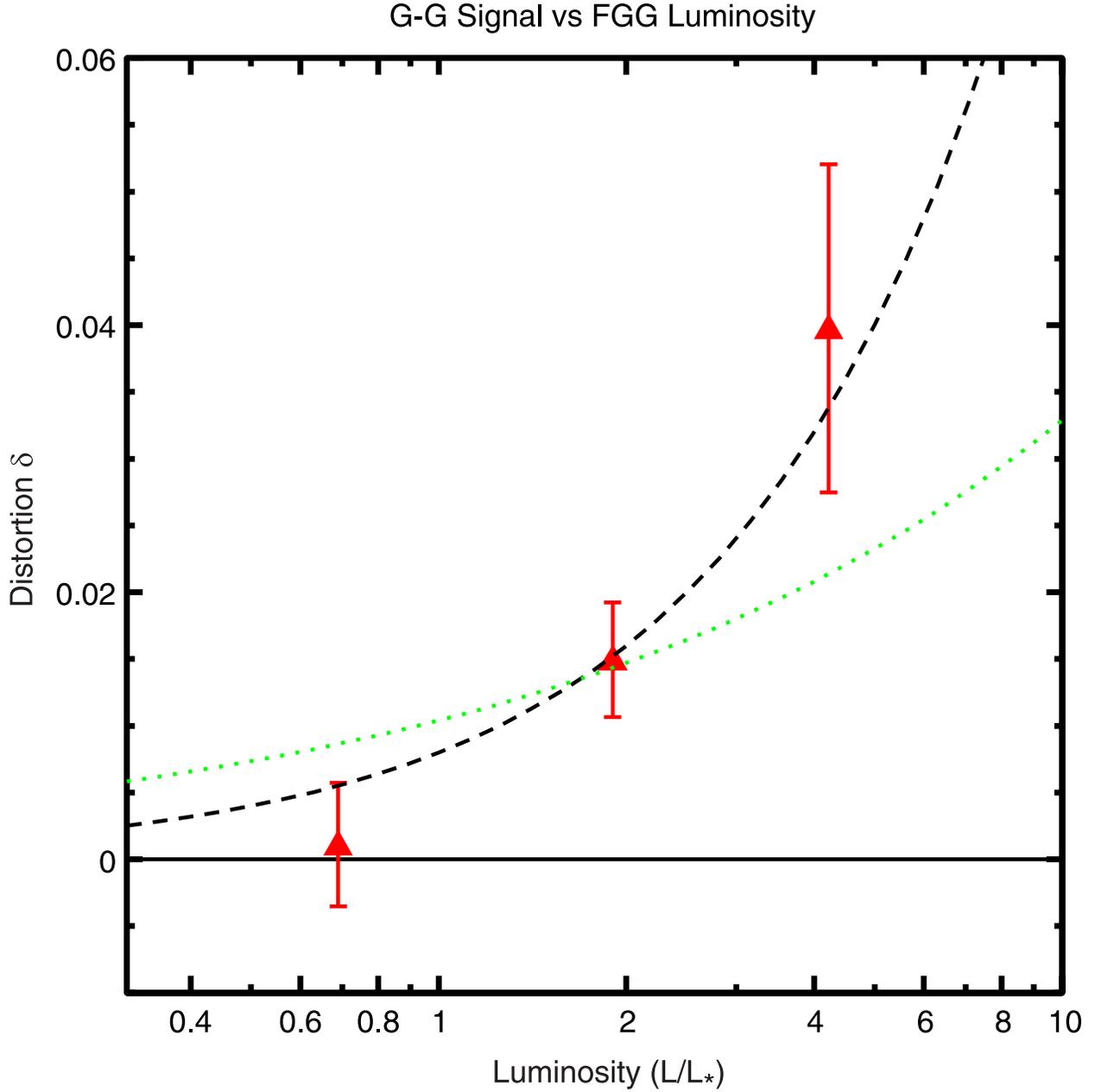}
\caption[dummy]{\small
The measured distortion $\delta$ (at $R=60\hkpc$) is plotted for three
bins of foreground galaxy luminosity $L$.  Dependence of distortion (and
hence mass) upon $L$ is clearly detected.  The dashed line is the best-fit
mass-traces-light model ($M\propto L$), which fits the data well;
the dotted line is the best-fit Faber-Jackson model ($M\propto
L^{0.5}$), which is only marginally acceptable.
\label{sigvsL}
}
\end{figure}

The largest uncertainty in our calculation of $\Omega$ is the
exponent $\beta$ describing dependence of galaxy mass upon
luminosity.  This parameter is normally assumed to be near $0.5$ based
upon dynamical measures in the luminous parts of galaxies, but it is
purely an ansatz for mass at radii beyond 30\hkpc.  We have divided
our foreground sample into 3 bins of luminosity and fit an isothermal
distortion profile to an annulus $15\hkpc < R < 240\hkpc$ for each bin.
In Figure~\ref{sigvsL} the dependence of
distortion (mass) upon $L$ is readily apparent.  This is in contrast
to the \citet{Za97} satellite galaxy study, which did not
see a significant correlation between luminosity and halo mass at
200\hkpc.  The galaxies certainly ``know'' about the density of the
halos in which they are embedded.

The dependence of $\delta$ upon $L$ is readily fit by the
Equations~(\ref{vcbeta}) for the mass-traces-light value of
$\beta=1$, ($\chi^2/\nu = 1.16/2$).  The Faber-Jackson value of
$\beta=0.5$ is somewhat disfavored ($\chi^2/\nu=4.93/2$, $Q=0.09$).
Treating $\beta$ as a free parameter in a fit to the three bins yields
a 95\% CL range of $0.6<\beta<2.4$.
This result is intriguing and demonstrates the power of weak lensing
to elucidate the relation between luminous and dark matter.
The traditional $\beta=0.5$ model is not strongly excluded, however,
so we will await further data before drawing conclusions.

A similar marginal difference is seen between the masses of
absorption-spectrum and emission-spectrum galaxies in the foreground
sample, but meaningful results must await better statistics.

\section{Comparison to Other Determinations}
\subsection{Comparison with Other Galaxy-Galaxy Lensing Estimates}
We place our result in context first with previous measures of
galaxy-galaxy lensing.  Most previous works were parameterized by the
halo circular velocity of an $L_\ast$ galaxy in the isothermal portion
of the halo, which we have called $v_\ast$.  Our measure of
$\delta_\ast$ gives 1-sigma limits on $v_\ast$ of
\begin{equation}
\label{vast}
v_\ast = \cases{
176\pm22\,{\rm km\,s}^{-1} & $(\beta=0.5)$ \cr
164\pm20\,{\rm km\,s}^{-1} & $(\beta=1.0)$.\cr
}
\end{equation}
This is in reasonable agreement with the determinations of \citet{Br96}
($v_\ast=220\pm80\,{\rm km\,s}^{-1}$, 90\% CL), 
\citet{Hu98} ($v_\ast=210\pm40\,{\rm km\,s}^{-1}$) and of F00 (200--280
\kms, 95\% CL), given that the disparate definitions of $L_\ast$
have been used (the LCRS value is on the low side).  
We emphasize, however, that ours is the first
determination of this quantity to actually measure the luminosities of
the foreground galaxies to give a result which is not heavily
dependent upon assumptions about the luminosity and redshift
distributions of the galaxy populations.

\subsection{Comparison with Other Dynamical Estimates}
\citet{Za94} use the satellite galaxy data to constrain
the mass $M_{200}$ enclosed within a 200\hkpc\ sphere of their typical
spiral to be 1.1--2.0$\times10^{12}h^{-1}\;M_\odot$.  Our results imply a mass
within 200\hkpc\ of
$(1.25\pm0.25)(L/L_\ast)\times10^{12}h^{-1}\;M_\odot$ for $\beta=1$,
or $(1.44\pm0.36)\sqrt{L/L_\ast}\times10^{12}h^{-1}\;M_\odot$ for
$\beta=0.5$.  There is general agreement, but a more specific comparison
is not possible because Zaritsky \& White find no dependence of mass
upon primary luminosity ($\beta\approx0$ in our terminology) while we
find a relatively strong dependence ($\beta\ge 0.5$).

\subsection{Comparison to the Tully-Fisher Relation}
It is of interest to compare the circular velocity of the halo to the
circular velocity of the gaseous disk.
\citet{Tu98} present a calibration of the Tully-Fisher relation
(TFR) 
for Ursa Major and Pisces cluster spirals in the $R$ band.  The slope
of their $R$-band TFR is equivalent to $\beta=0.62$.  But
they also find that the extinction is stronger in brighter galaxies,
and the LCRS magnitudes are not corrected for extinction, so the
mass-luminosity relation in the LCRS data would appear somewhat
shallower, $\beta=0.58$, for a typical galaxy at $\sim60\arcdeg$
inclination.  For this value of $\beta$, the lensing data suggest
a circular velocity $v_\ast=176\pm22\,{\rm km\,s}^{-1}$ at an
effective radius of $\approx60\hkpc$.  The TFR calibration yields a
disk circular velocity of $160\,{\rm km\,s}^{-1}$ for a
typically-inclined spiral at the LCRS value of $L_\ast$.  This
consistency suggests that the isothermal mass profile observed at
$\sim10\hkpc$ in the disk matches fairly well onto the
halo profile at $\sim100\hkpc$.  Even if galaxies have perfect
isothermal profiles, we might expect the lensing $v_\ast$ to be
slightly higher than the TFR value (for late-type spirals) because of
the inclusion of early types in the lensing foreground sample.

\section{Conclusions}
The relations between galaxy luminosity and halo mass determined
herein provide basic constraints to models of galaxy formation and
evolution.  We find, perhaps surprisingly, that the halo circular
velocity at 60--100\hkpc\ radii are similar to the disk circular
velocities measured at $\lesssim10\hkpc$, extending
the ``disk-halo conspiracy'' to larger radii than previously known.
There are hints in the data, however, that this conspiracy breaks down
in the sense that halo masses vary more sharply with galaxy luminosity
then would be indicated by the Tully-Fisher or Faber-Jackson
relations.

The matter density $\Omega_m$ has been measured by many methods in the
past, and there is a currently fashionable consensus that
$\Omega_m\approx0.3$.  Our results (Equation~[\ref{omega}]) are in
agreement with this consensus given that our method produces
only a lower limit.  Our limits are, however, close to the consensus
value, especially if $\beta\approx0.5$.
Our lower limit on $\Omega$ is determined by a
direct inventory of the matter in the Universe; the only untested
assumption is that the formula for gravitational deflection of light
can be extrapolated from Solar-System scale (where it has been
verified) to galactic scales.
Other determinations of
$\Omega$ rely upon assumptions about the nature of our Universe.
Agreement between our direct inventory and these other measures serves
as valuable and unique verification that these assumptions---often
fundamental to our current view of cosmology---are correct.

The more precise estimates of $\Omega_m$ include the following:
Dynamical studies of galaxy clusters, combined with the assumption
that the $M/L$ ratio for clusters is universal, yield
$\Omega_m=0.24^{+0.05}_{-0.09}$ \citep{Ca96}.
Another cluster-based means to $\Omega$ is to measure the fraction 
$f_b$ of the cluster matter that is baryonic, assume that the cluster
$f_b$ is universal, and then use the baryon density
$\Omega_b$ from Big Bang nucleosynthesis to estimate
$\Omega_m=\Omega_b/f_b$.  \citet{Mo99} conclude that
$\Omega_m<0.30\pm0.04$ (for $h=0.7\pm0.1$).  

Other estimates of $\Omega$ measure its effects upon global geometry
or the growth rate of fluctuations in the Universe.  In the former
category, the combination of high-redshift supernovae measurements and
the location of the cosmic background Doppler peaks measured by
MAXIMA-1 constrain $0.25<\Omega_m<0.50$ at 95\% CL \citep{Ba00}.
Peculiar velocity studies determine the parameter
$\beta_I\approx\Omega^{0.6}/b$ to be $\approx0.5\pm0.05$ \citep{St98}, 
with $b$ a galaxy bias parameter.  Taking $b\approx1$
gives a value of $\Omega\approx0.3$, again slightly above our halo density.
Conversely our lower limit on $\Omega$ implies a bias parameter for
IRAS galaxies $b\gtrsim0.7$.

Thus our results may be taken as independent evidence that these
underlying assumptions are correct:  cluster $M/L$ and baryon
fractions can be taken as representative; general relativity does
properly describe the geometry of the Universe to $z\approx1000$ given
our directly measured matter content; and large-scale motions are
consistent with the predictions of gravitational instability.

Conversely, we may take the $\Omega_m\approx0.3$ value as truth and
conclude from these lensing results that most of the matter in the
Universe is indeed contained within 260\hkpc\ of normal galaxies.
This implies that galaxies missed by the LCRS survey---{\it e.g.}
low-surface-brightness galaxies or exotic dark galaxies---cannot be
the dominant reservoirs of mass in the Universe.  Likewise there
cannot be a dominant matter component which is not clustered around
galaxies.  Most of the mass in the Universe appears to have been
located, though of course its material nature remains unknown.

The accuracy of our results is currently limited by the number of
foreground/background pairs in our survey.  This will be remedied in
dramatic fashion by the full Sloan Digital Sky Survey, which will
obtain imaging to $r^\prime<22$ around roughly {\em three orders of
magnitude} more foreground galaxies than used in this study with a
concomitant increase in the number of pairs.  This will allow for
non-parametric mapping of the shear profile as a function of radius,
galaxy luminosity, and galaxy type, with much lower noise than in 
our LCRS lensing survey.  This weak lensing information will be unique
in unveiling the relation between the various classes of galaxies and
the dark halos in which they reside.
With the random errors reduced, closer
attention will have to be paid to some of the $\lesssim5\%$
corrections we have ignored (as in F00), but these are tractable.

Our estimates of $M_\ast$ and $\Omega$ are lower bounds since the
profile of the halo has not yet been determined beyond 260\hkpc.
Beyond this radius it becomes inappropriate to speak of ``the halo''
of a galaxy, as the mass profile of a given galaxy becomes 
inseparable from the mass associated with neighboring galaxies.
More stringent limits on $\Omega$ will thus require a more elaborate
formalism for describing the effects of galaxy correlations upon the
weak lensing measurements.

\acknowledgements
This work was supported by grant AST-9624592 from the National Science
Foundation.  We gratefully acknowledge the assistance of J. A. Tyson
and David Wittman with the construction and operation of the BTC
camera and their contributions to the data pipeline; Tim McKay and
Erin Sheldon for
continued discussions of SDSS lensing results; and the staff of CTIO
for their enthusiastic support of the BTC observations.


\begin{thebibliography}{}

\bibitem[Balbi \etal (2000)]{Ba00}
Balbi, A. \etal 2000, astro-ph/0005124

\bibitem[Bernstein \etal (2000)]{Be00}
Bernstein, G. M., Jarvis, R. M., Fischer, P., \& Smith, D. R., 2000
(in preparation) 

\bibitem[Bertin \& Arnouts (1996)]{Be96} 
Bertin, E., \& Arnouts, S. 1996 \aap, 117, 393

\bibitem[Brainerd, Blandford, \& Smail (1996)]{Br96} 
Brainerd, T.G., Blandford, R.D., \& Smail, I. 1996, \apj, 466, 623 

\bibitem[Carlberg \etal (1996)]{Ca96}
Carlberg, R. G. \etal 1996, \apj, 462, 32

\bibitem[Cohen \etal (2000)]{Co00} Cohen, J. G. et al. 2000, \apj\ (in
press) astro-ph/9912048

\bibitem[Crampton \etal (1995)]{Cr95}
Crampton, D., Le F\'evre, O., Lilly, S.J., \& Hammer, F. 1995, \apj, 455, 96

\bibitem[Dell'Antonio \& Tyson (1996)]{DA96}
Dell'Antonio, I.P. \& Tyson, J.A. 1996, \apjl, 473, L17

\bibitem[Fischer \& Tyson (1997)]{Fi97}
Fischer, P. \& Tyson, J. A. 1997, \aj, 114, 14

\bibitem[Fischer \etal (2000)]{F00}
Fischer, P. \etal 2000 , \apj\ (in press) astro-ph/9912119 (F00)

\bibitem[Griffiths, Casertano, \& Ratnatunga (1996)]{Gr96}
Griffiths, R.E., Casertano, S., Im, M., \& Ratnatunga, K.U. 1996,
\mnras, 282, 1195

\bibitem[Hudson \etal (1998)]{Hu98}
Hudson, M.J., Gwyn, S.D.J., Dahle, H., \& Kaiser, N., 1998, ApJ, 503, 531 

\bibitem[Kaiser, Squires, \& Broadhurst (1995)]{Ka95}
Kaiser, N., Squires, G., \& Broadhurst, T. 1995, \apj, 449 460

\bibitem[Landolt (1992)]{La92}
Landolt, A.U. 1992, \aj, 104, 340

\bibitem[Lin \etal (1996)]{Li96}
Lin, H., Kirshner, R. P., Shectman, S.A., Landy, S. D., Oemler, A.,
Tucker, D. L., \& Schechter, P. L. 1996, \apj, 464, 60

\bibitem[Mohr, Mathiesen, \& Evrard (1999)]{Mo99}
Mohr, J. J., Mathiesen, B., \& Evrard, A. 1999, \apj, 517, 627

\bibitem[Shectman \etal (1996)]{Sc96}
Shectman, S. A., Landy, S. D., Oemler, A., Tucker, D. L., Lin, H., Kirshner, R.P., \& Schechter, P. L. 1996, \apj, 470, 172

\bibitem[Smith (2000)]{Sm00}
Smith, D. R. 2000, PhD Thesis, University of Michigan

\bibitem[Stebbins, McKay, \& Frieman (1996)]{St96}
Stebbins, A., McKay, T., \& Frieman, J. 1996, in ``IAU 173:
Astrophysical Applications of Gravitational Lensing,''
eds. C. S. Kochanek \& J. N. Hewitt (Kluwer), 75

\bibitem[Strauss \& Willick (1998)]{St98}
Strauss, M. \& Willick, S. 1998, \apj, 507, 64

\bibitem[Tully \etal (1998) ]{Tu98}
Tully, R. B. \etal 1998, \aj, 115, 2264

\bibitem[Tyson (1988)]{Ty88}
Tyson, J.A., 1988, \aj, 96, 1

\bibitem[Tyson \etal (1984)]{Ty84}
Tyson, J.A, Valdes, F., Jarvis, J.F., \& Mills, A.P. 1984, \apjl, 281, L59

\bibitem[Wittman \etal (1998)]{Wi98}
Wittman, D.M., Tyson, J.A., Dell'Antonio, I.P., Bernstein, G.M., \&
Smith, D.R. 1998,  \procspie 3355, 626-634

\bibitem[Zaritsky \& White (1994)]{Za94}
Zaritsky, D. \& White, S.D.M., 1994, \apj, 435, 599
 
\bibitem[Zaritsky \etal (1997)]{Za97}
Zaritsky, D., Smith, R., Frenk, C., \& White, S.D.M., 1997, \apj, 478, 39

\end{thebibliography}
\end{document}